\documentclass[article]{aastex63}
\usepackage{float}
\usepackage{amsmath}
\usepackage{comment}
\usepackage{graphicx}
\usepackage[flushleft]{threeparttable}
\usepackage{changepage}
\usepackage{sidecap}
\sidecaptionvpos{figure}{c}

\usepackage{kpfonts}
\usepackage[caption=false]{subfig}
\begin{document}

\title{Minimizing Star Spot Contamination of Exoplanet Transit Spectroscopy Using Alternate Normalization}

\shorttitle{Minimizing Star Spot Contamination}
\shortauthors{Deming et al.}

\correspondingauthor{Drake Deming}
\email{ldeming@umd.edu}

\author[0000-0001-5727-4094]{Drake Deming}
\affiliation{Department of Astronomy, University of Maryland,
   College Park, MD 20742, USA}
\affiliation{NASA's Nexus for Exoplanet System Science, Virtual Planetary Laboratory Team, Box 351580, University of Washington, Seattle, WA 98195, USA}  

\author[0000-0003-3429-4142]{Miles H. Currie}
\affiliation{NASA Goddard Space Flight Center, Greenbelt, MD 20771, USA}
\affiliation{NASA's Nexus for Exoplanet System Science, Virtual Planetary Laboratory Team, Box 351580, University of Washington, Seattle, WA 98195, USA}

\author[0000-0002-1386-1710]{Victoria S. Meadows}
\affiliation{Department of Astronomy and Astrobiology Program, University of Washington, Box 351580, Seattle, WA 98195, USA}
\affiliation{NASA's Nexus for Exoplanet System Science, Virtual Planetary Laboratory Team, Box 351580, University of Washington, Seattle, WA 98195, USA}

\author[0000-0002-1046-025X]{Sarah Peacock}
\affiliation{NASA Goddard Space Flight Center, Greenbelt, MD 20771, USA}
\affiliation{University of Maryland, Baltimore County, Baltimore, MD, 21250, USA}

\begin{abstract}
Recently, Currie et al. simulated the detection of molecules in the atmospheres of temperate rocky exoplanets transiting nearby M-dwarf stars. They simulated detections via spectral cross-correlation applied to high resolution optical and near-IR transit spectroscopy using the ELTs.  Currie et al. did not consider the effect of unocculted star spots,  but we do that here for possible detections of molecular oxygen, carbon dioxide, methane, and water vapor.  We find that confusion noise from unocculted star spots becomes significant for large programs that stack tens to hundreds of transits to detect these molecules. Noise from star spots increases with greater spot filling factors, and star spot temperature has less effect than filling factor.  Nevertheless, molecular oxygen, carbon dioxide, and methane could be detected in temperate rocky planets transiting nearby M-dwarfs without correcting for star spots. Water vapor detections are the most affected, with star spots contaminating the exoplanet signal as well as producing extra noise. Unocculted spots only affect transit spectroscopy when normalizing by dividing by the total flux from the star.  We describe an alternate normalization method that minimizes star spot effects by deriving and implementing an unspotted proxy spectrum for the normalization.  We show that the method works in principle using realistic levels of random observational noise. Alternate normalization would be broadly applicable to all types of transit spectroscopy, and we discuss challenges to applying it in practice.  We also outline a comprehensive approach that has the potential to overcome those challenges.

\end{abstract}

\vspace{5mm}

\section{Introduction}

The new generation of Extremely Large Telescopes (ELTs) will have important applications for studies of exoplanets, both directly imaged planets and transiting planets \citep{konopacky_2023}.  Recently, \citet{currie_2023} and \citet{hardegree_2023} studied the sensitivity of the ELTs for transit spectroscopic detection of molecules in the atmospheres of temperate rocky planets that transit M-dwarf stars.  

M-dwarf stars have convective envelopes that can generate strong photospheric magnetic fields \citep{henry_2024, west_2008}.  The temperature perturbations associated with magnetic activity can interfere with detections of molecules in the transiting exoplanetary atmosphere \citep{rackham_2018}. In the worst case, planetary molecules could also be present in cool star spots, and could produce false detections \citep{wakeford_2019b}.  Unocculted spots are potentially the biggest problem because, unlike spots crossed by the planet, they produce no direct signatures in the transit light curves.  The cross-correlation (hereafter, CC) detection technique studied by \citet{currie_2023} has great sensitivity to exoplanetary molecules because it integrates over all of the wavelengths within the span of a given spectrum from an ELT spectrometer.  \citet{currie_2023} focused exclusively on that sensitivity, and left the evaluation of star spot effects to a future paper.

\vspace{5mm}
\subsection{Motivation and Organization of This Paper}

In this work, we expand on the calculations done by \citet{currie_2023} to include the effect of unocculted star spots, and we propose an observational and data analysis method that minimizes contamination by those spots.  \S\,\ref{sec: cc} explains the principle of our molecular detections, and benchmarks our results versus \citet{currie_2023}.  \S\,\ref{sec: spots} explores the potential effect of unocculted star spots, including how they increase the noise and contaminate detections of exoplanetary molecules. \S\,\ref{sec: alternate} introduces our alternate normalization method to produce exoplanetary transit spectra that minimize noise and contamination from unocculted star spots.  \S\,\ref{sec: caveats} lists some challenges that must be overcome, and \S\,\ref{sec: comprehensive} discusses a comprehensive approach to deal with some of those challenges. \S\,\ref{sec: summary} summarizes our results. 

\section{Cross-correlations}\label{sec: cc}

\begin{SCfigure}
\centering
\includegraphics[width=4in]{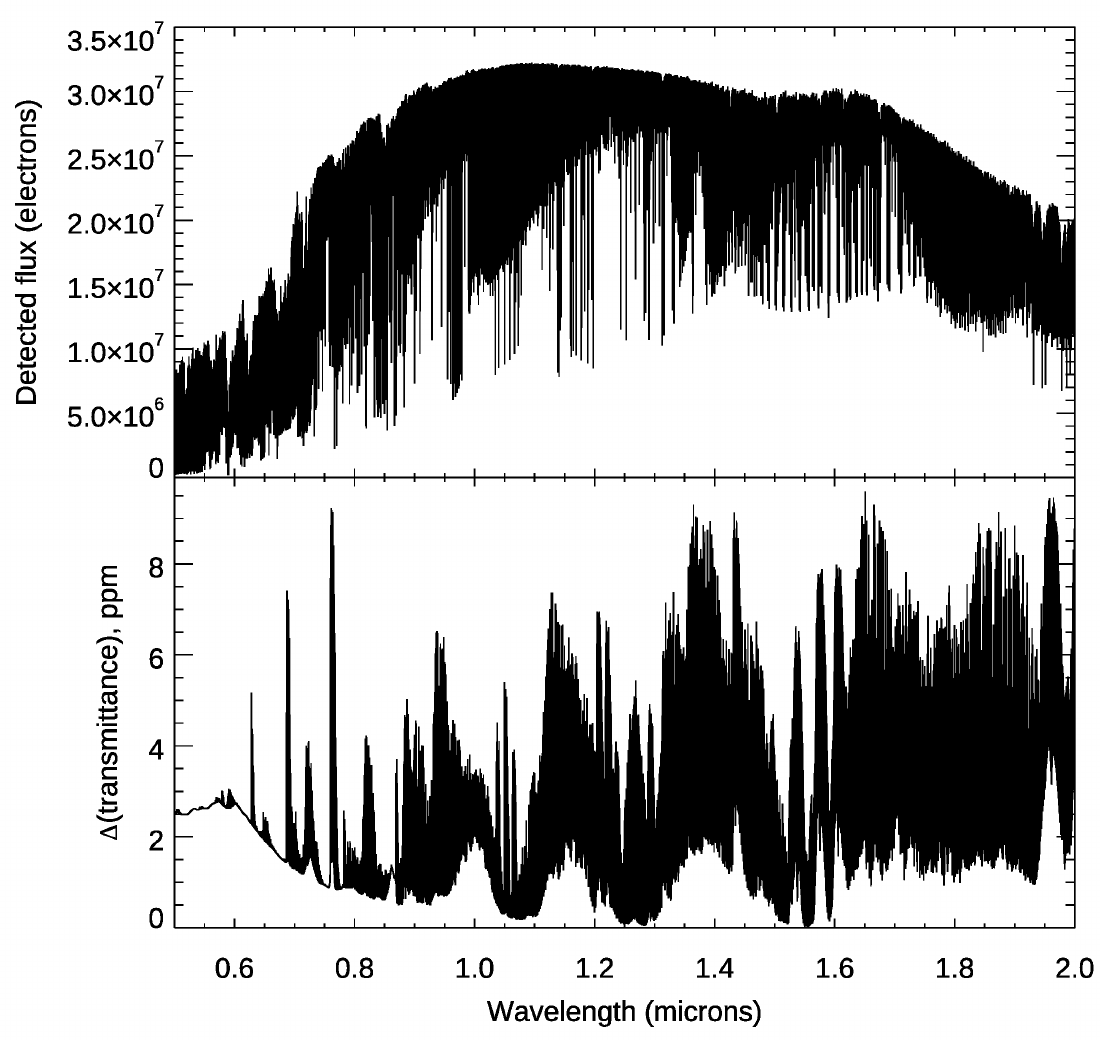}
\caption{{\it Upper panel:} Electrons detected by an ELT high-resolution spectrometer ($R=500,000$) integrated over the transit of a temperate planet orbiting an M3V star at 12 parsecs distance (Table~\ref{tab: planets}).  {\it Lower panel:} Transmission spectrum of the planet (in parts-per-million, minus the minimum transmission value). \label{fig: spectra_fig} }
\end{SCfigure}

The atmospheric signals due to rocky exoplanets in transit are small (Figures~\ref{fig: spectra_fig} \& \ref{fig: trans_template}), and even the ELTs will not collect sufficient photons to enable detection of individual vibration-rotation lines in molecular bands. \citet{currie_2023} calculated the photon flux from M-dwarf stars potentially hosting rocky planets, using models from \citet{peacock_2019a, peacock_2019b, peacock_2020}, as measured by the ELTs.  Figure~\ref{fig: spectra_fig} shows the number of electrons detected per resolution element ($R=500,000$) for one of those modeled M-dwarfs (M3V) during a full transit of a temperate planet, together with the transmission of the exoplanetary atmosphere.  Although the transmission per line is very small ($< 10$\,ppm) the CC technique efficiently averages over all of the line structure in a band, and makes the detection possible, albeit still difficult.  These detections require averaging tens, or even hundreds, of transits \citep{snellen_2013, lopez-morales_2019, hardegree_2023, currie_2023}.  Nevertheless, the ELTs are expected to be long-term assets for ground-based astronomy, and it is reasonable to suppose that they could support long-term programs to characterize the atmospheres of rocky exoplanets. 

CC detections of exoplanetary molecules can be made for non-transiting planets \citep{brogi_2014}, as well as transiting planets.  In either case, data taken over a full orbit are desirable (the more data, the better), but it is also possible to use CCs for data that are concentrated only near transit \citep{esparza-Borges_2023}.  When full orbit data are used, the CC can exploit the Doppler shift of the exoplanet's spectrum to discriminate the exoplanetary signal from the noise \citep{deKok_2014}.  When the data are concentrated near transit, the Doppler effects are less useful, but the CC nevertheless can detect exoplanetary molecules by the weighting and averaging over wavelength.  That near-transit case in what we consider here, following \citet{currie_2023}.

\subsection{Transit formulae for cross-correlations}\label{sec: formulae}

We describe our methodology by first giving the formulae that govern our simulated CC detections.  These formulae are well known, but we reiterate them here to frame our discussion.  We consider a star with both quiet photospheric regions, and star spots. 
For simplicity, we ignore other active phenomena such as faculae, but our methodology could in principle be expanded to include facular regions. The in-transit flux observed at wavelength $\lambda$ is:
\begin{equation}
F_{in}^{\lambda} = F_{q}^{\lambda}(1-(R_{p}^{\lambda}/R_{s})^2) + F_{sp}^{\lambda},
\end{equation}
where $F_q$ is the flux from the quiet (non-spotted) portions of the stellar disk facing the Earth, and $F_{sp}$ is the flux from the regions of the stellar disk that contain unocculted star spots. $R_{p}^{\lambda}$ is the wavelength-dependent radius of the planet, and $R_{s}$ is the equivalent radius of the star.  It is equivalent in the sense of being the radius corresponding to the area of the quiet regions of the star, albeit the total geometric radius could also be used to a good approximation. 

The out-of-transit flux is:
\begin{equation}
F_{out}^{\lambda} = F_{q}^{\lambda} + F_{sp}^{\lambda},
\end{equation}
Assuming that the unocculted star spots do not vary on the time scale of a transit, the signal we process in our CC analysis is given as:

\begin{equation}
S_{\lambda} = ( F_{out}^{\lambda} - F_{in}^{\lambda} )/F_{out}^{\lambda},
\end{equation}

and that breaks down as:

\begin{equation}
S_{\lambda} = F_{q}^{\lambda}(R_p^{\lambda}/R_s)^2/(F_{q}^{\lambda} + F_{sp}^{\lambda})
\end{equation}

The CC itself is a function of a lag in wavelength ($\delta\lambda$), and at each wavelength ${\lambda}$ the signal $S_{\lambda}$ is multiplied by a template function ($T_{\lambda + \delta\lambda}$), and the product is integrated over wavelength:

\begin{equation}
  CC_{\delta\lambda} = \int A_{\lambda} S_{\lambda} T_{\lambda + \delta\lambda} \,d{\lambda},
\end{equation}

where the integral extends over the range in wavelength that encompasses the molecular absorption of interest. $A_{\lambda}$ is an apodizing function that attenuates edge effects; we use a cosine bell \citep{bracewell_2000} applied to the highest and lowest 10\% in wavelength.  Our apodization and calculation of the cross-correlation function and its signal-to-noise ratio (described below) are independent of \citet{currie_2023}, but are equivalent in the results (\S\ref{sec: comparison}).

\subsubsection{Limb Darkening}\label{sec: limbdark}
Eq.(1) implicitly ignores stellar limb darkening, because of the way the $R_{p}^{\lambda}/R_{s}$ term is used.  We choose to ignore limb darkening in order to present the effect of unocculted star spots with maximum clarity. Limb darkening is important for analyzing transit spectroscopy by fitting modeled transit curves to transit time series data, but the CC principles and methods we articulate here are not very sensitive to limb darkening, as we now illustrate with an example. A star spot on a limb darkened star typically blocks a larger fraction of the stellar flux than it would in the absence of limb darkening. The difference is maximum when the spot is centered on the star, because the stellar intensity is maximum there.  Using 4-parameter limb darkening\footnote{https://exoctk.stsci.edu/limb-darkening}, we calculated the fraction of stellar flux blocked in the limb-darkened case versus the case of a uniform stellar disk at a wavelength of 1.5\,$\mu$m.  We considered a star spot having an area equal to 10\% of the projected disk of Proxima Centauri (Table~\ref{tab: planets}).  Centering that spot on the stellar disk, we find that the 10\% blockage of a uniform star becomes 11.8\% on the limb-darkened star, i.e. approximately 18\% greater.  Other stars and wavelengths give similar results. In \S\ref{sec: spot_noise} we consider projected areas of star spots that vary by two orders of magnitude (from 0.2\% to 20\%).  Hence the effect of limb darkening is not significant compared to the range of the free parameter that we use to explore star spot effects.

\subsection{Calculation of the Cross-correlations}\label{sec: calc}

We calculate the CC functions at high spectral resolution using a custom code in IDL.  The PHOENIX models have $R=5.0 \times 10^5$ \citep{husser_2013}, hence the limiting resolving power is $R=1.0\times 10^5$ used by \citet{currie_2023} to calculate the transit spectra.  We convolve our template and the PHOENIX models to closely match that resolving power, and we interpolate the product in Eq.(5) onto a numerical integration grid that is approximately 40 times finer than the limiting resolution.

The modeled transmission function in the exoplanetary atmosphere ($R_p^2/R_s^2$ versus $\lambda$, producing $S_{\lambda}$) increases with greater exoplanetary absorption, whereas absorption in star spots (e.g., by water vapor) causes the flux to decrease. The modeled planet and star spots would thereby produce CC peaks of opposite sign. To reconcile those opposite signs, we flip the sign of the modeled exoplanetary transmission values.  Consequently, molecular absorption in both  the exoplanet and star spots is detected by negative peaks in the CC function, similar to an absorption line.  We emphasize that the negative peaks in our case indicate correlation, not anti-correlation.

\subsection{Models \& planets}
For our models of exoplanetary transmission in transit we use the pre-industrial Earth models (modern Earth with no anthropogenic surface fluxes) of \citet{currie_2023}, and consider only the molecules carbon dioxide, methane, molecular oxygen and water vapor. That focus is appropriate because our goal is to evaluate the effect of star spots, not to explore the detectability of diverse exoplanetary atmospheres.  The planets we model are hypothetical, but the stellar hosts are based on real stars (M2V to M8V) at hypothetical distances from Earth, as per \citet{currie_2023}. Following \citet{currie_2023} we place the planetary systems at either 5- or 12\,parsecs from Earth. The properties of the star/planet systems are given in Table~\ref{tab: planets}.

We use high-resolution ($R=500,000$) PHOENIX stellar models from \citet{husser_2013}, and also as calculated by \citet{peacock_2019a, peacock_2019b, peacock_2020}.  We use those models to represent both the quiet photospheres and the star spots, adopting a temperature contrast (${\Delta}T$) from spot to stellar photosphere.  In that respect, we note that MHD models of stellar active regions indicate that their spectra are not optimally represented by a non-magnetic stellar photosphere having a different temperature \citep{smitha_2025}. Our evaluation of star spot effects should be re-examined when the MHD models are more developed and widely distributed.  However, in Section~\ref{sec: alternate} we show how the contamination from unocculted star spots can be minimized, and our method is independent of the spectra of the star spots.

\begin{table} [h] 
\begin{tabular}{llllll}
\small
System &  Spectral type & T$_{star}$ (K) & $\frac{R_{star}}{R_{\odot}}$  &  a$_{orb}$ (AU) & T$_{planet}$ (K)  \\
  \hline
  GJ\,832          & M2V & 3539 &  0.499 &  0.24  & 247  \\
  GJ\,436          & M3V & 3416 &  0.464 &  0.19  & 258   \\
  GJ\,876          & M4V & 3201 &  0.376 &  0.16  & 237  \\
  Proxima Centauri & M6V & 2992 &  0.141 &  0.041 & 251 \\
  TRAPPIST-1       & M8V & 2566 &  0.114 &  0.027 & 255 \\
\end{tabular} 
\caption{\small Planetary systems studied in this work. We use the host stars listed here to represent hypothetical hosts at different distances from Earth. The stellar temperatures are from \citet{pineda_2021}, except GJ\,436 that is from TEPCat \citep{southworth_2011, southworth_2012}, and TRAPPIST-1 that is from \citet{agol_2021}. The other columns are from \citet{currie_2023}, except T$_{planet}$ that is calculated from our adopted stellar temperatures, and using an albedo of zero and redistribution of stellar irradiance over the entire planet. }
\label{tab: planets}
\end{table}

\subsubsection{Templates}

\begin{SCfigure}
\centering
\includegraphics[width=4in]{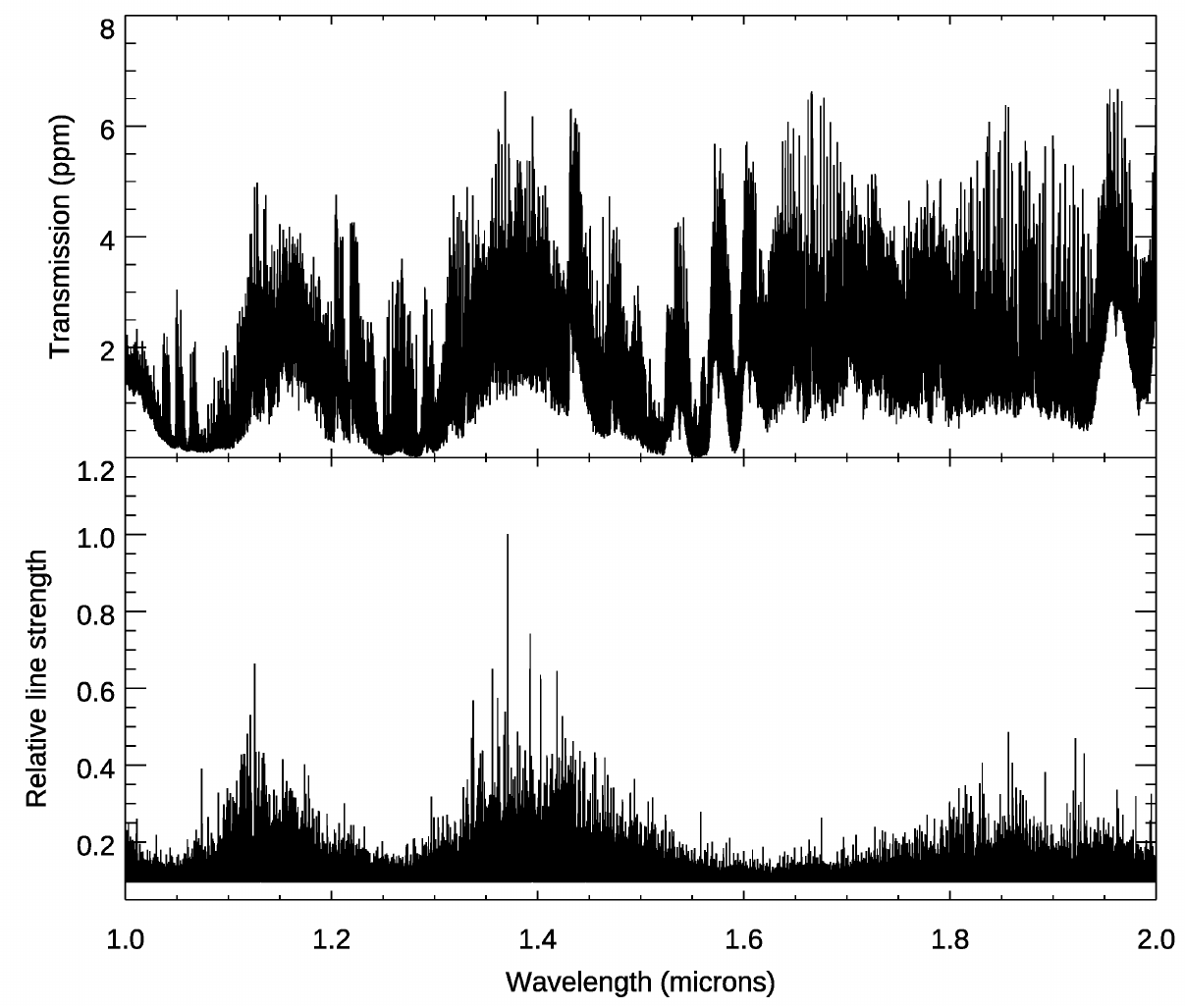}
\caption{{\it Upper panel:} Transit spectroscopy signal $R_p^2/R_s^2$ (minus the minimum value, and given in parts-per-million) for a temperate planet orbiting an M2V star, and adopting the ''pre-industrial Earth" planetary model from \citet{currie_2023}.  {\it Lower panel:} Template of line strengths for water vapor used in our cross-correlations. \label{fig: trans_template} }
\end{SCfigure}

\citet{currie_2023} used the planet's transmission model itself as a CC template, but we have taken a more agnostic viewpoint.  When calculating CCs for real observed data, we will probably not know the physical state of the exoplanetary atmosphere beyond an estimate of the equilibrium temperature.  We therefore construct our templates (one template per molecule) using line strengths from HITRAN \citep{rothman_1998, gordon_2022}. Each template is a collection of $\delta$-functions at wavelengths of molecular lines, with the amplitude of each $\delta$-function equal to the log of the line strength, after normalizing the strengths to unity for the strongest line in the wavelength range (0.5- to 2\,$\mu$m).  Because real planetary molecular lines are not $\delta$-functions, we convolve the template with a Gaussian having a FWHM of 2.7 km/sec, to simulate Doppler and pressure broadening in the exoplanetary atmosphere. We determined the width of the Gaussian as the value that maximized the amplitude of the CC functions in a high SN case (carbon dioxide in an M3V system).  We deem that to be a realistic adjustment, because observers could optimize the signal-to-noise ratio (S/N) of their detections in a similar manner.  We also tried scaling the relative line strengths of the molecular lines in the template as a function of the assumed excitation temperature, and we found a broad maximum in CC amplitude for excitation temperatures between 250- and 300\,Kelvins.  Because it was a broad maximum, we decided to use the HITRAN line strengths (for 296\,K) without scaling.  Figure~\ref{fig: trans_template} shows the exoplanetary transmission for a temperate planet orbiting an M2V star, compared to our template for water vapor.

\subsubsection{Telluric absorption}

There is significant telluric absorption in our wavelength range. Some planetary lines will overlap telluric lines, and the telluric absorption will reduce the CC signal, especially for water vapor.  However, the radial velocities of the stars relative to Earth will reduce that overlap. Following \citet{currie_2023}, we adopt a radial velocity (RV) relative to the observer of 22 km/sec for each planet (but for real observations, that RV will vary due to the 30 km/sec orbital velocity of the Earth).  Our telluric spectrum is from SkyCalc\footnote{https://www.eso.org/observing/etc/bin/gen/form?INS.MODE=swspectr+INS.NAME=SKYCALC} for a zenith distance of 30-degrees and using 2.5 millimeters of precipitable water vapor.  Ground-based observers can correct for telluric absorption to a large degree using early-type comparison stars (that have few stellar absorption lines).  However, that does not compensate for the greater noise at the wavelengths of telluric lines due to fewer photons.  Hence, we account for the loss of telluric-absorbed photons at high resolution when calculating the stellar photon noise, but we assume that the telluric absorption effect on line profiles has been removed from the CC calculation. 

\subsubsection{Example of a cross-correlation calculation}

To determine the S/N of a CC detection for a given molecule, we do two cross-correlations: one for the simulated planet plus photon noise, and one for a pure photon noise signal that does not contain structure due to the shape of $S_{\lambda}$.  The S/N of the detection is the amplitude of the CC dip divided by the standard deviation of the noise CC.  The noise signal is dominated by the total photon noise (versus $\lambda$) during the duration of a transit, and we adopt that the noise (relative to the stellar flux) decreases as the square root of the number of transits.  To remain consistent with \citet{currie_2023}, we include only the photon noise during transit, not additional noise from out of transit baseline observations.  Real observations may therefore obtain slightly degraded results compared to our calculations, depending on how many observations they acquire out of transit.

In addition to photon noise, there are other observational noise sources (e.g., zodiacal emission) discussed by \cite{currie_2023}.  With one exception, those noise sources are approximately an order of magnitude less than the photon noise.  Because independent noise sources add in quadrature, those minor sources of noise are not significant, and we ignore them.  The exception is airglow from hydroxyl (OH) lines that are very bright in the near-infrared \citep{meinel_1950, davies_2007}. In our CC calculations, we mask the wavelengths of many OH lines, so that they do not appear in the $S_{\lambda}$ term in Eq.(5).  Specifically, we mask all lines whose intensities are $\geq{0.001}$ times as strong as the brightest OH lines.  Our mask has a width of $\pm2$ Doppler widths for OH at the terrestrial mesopause, where $T \sim 200$\,K \citep{marsh_2006}. 
 
\begin{SCfigure}
\centering
\includegraphics[width=4in]{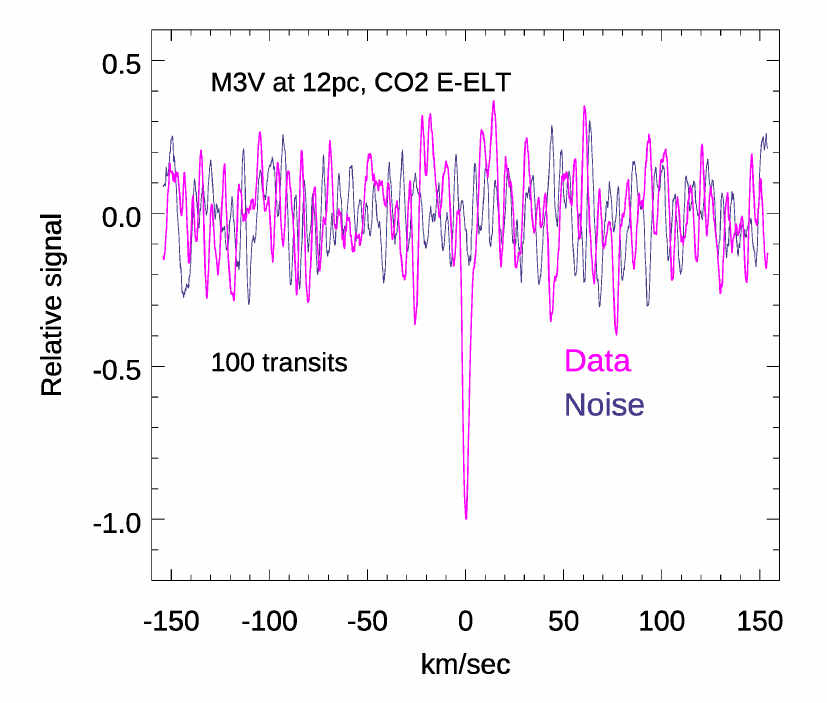}
\caption{Example of our calculated cross-correlation functions for carbon dioxide, that is clearly detected in the transit spectrum of a temperate planet orbiting an M3V star 12 parsecs from Earth, using 100 transits.  Two CCs are illustrated: the CC that models synthetic data including the planet transmission signal, and a CC from a pure noise source. For the latter CC, we generate noise at the same amplitude as in the simulated data.  The negative peak on the CC indicates a detection of exoplanetary absorption, see \S\ref{sec: calc} for an explanation of our sign convention.  The S/N is 8 for this simulated detection. \label{fig: cc_example} }
\end{SCfigure}

\citet{currie_2023} restricted the range of the CC-integration to specific vibration-rotation bands of each molecule.  They wanted to inform future observers and instrument builders as to what bands are the most detectable.  In contrast, we use the entire 0.5- to 2\,$\mu$m region because we want to optimize the total S/N so that star spot effects can be defined with the greatest clarity. Figure~\ref{fig: cc_example} shows our CC calculation for a high-S/N detection of carbon dioxide in 100 transits of a temperate planet orbiting an M3V at 12 parsecs distance.

\subsection{Comparison to Currie et al.}\label{sec: comparison}

\begin{SCfigure}
\includegraphics[width=0.35\textwidth]{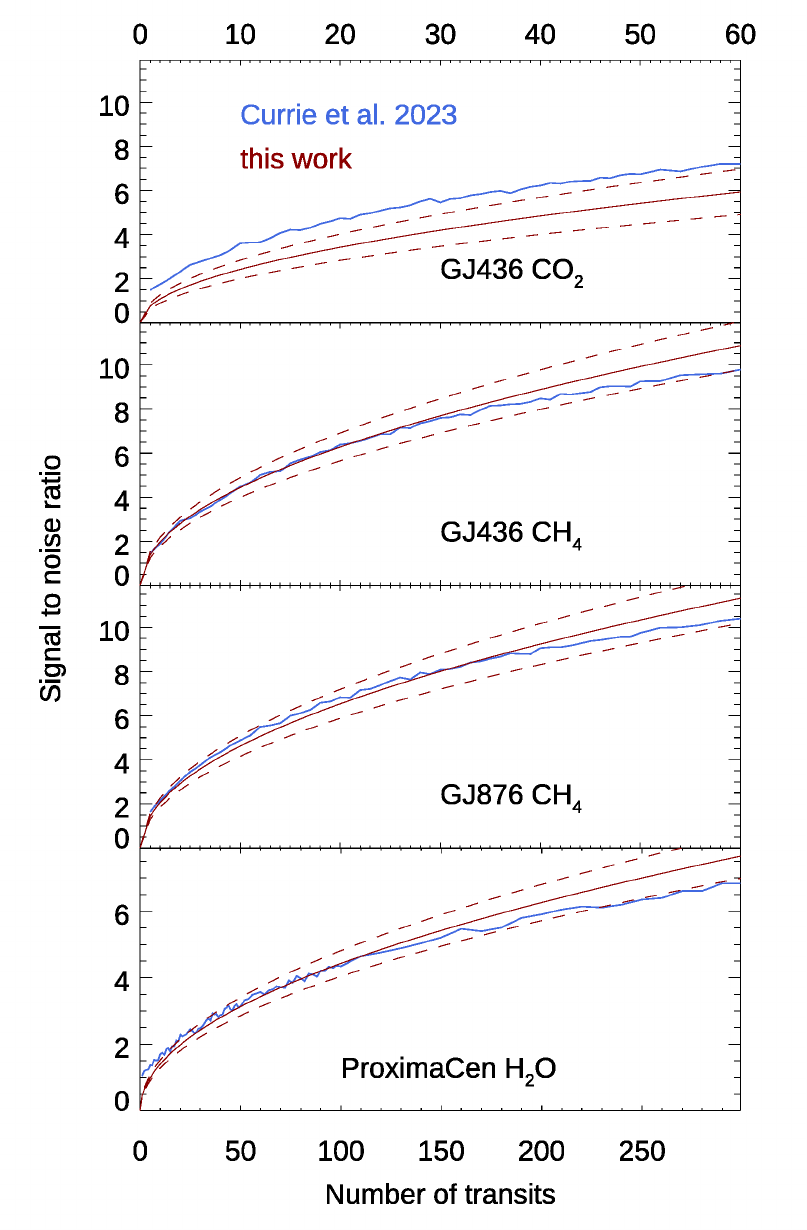}
\caption{Comparison of our calculated S/N ratios for cross-correlation detections with those of Currie et al., showing four different planetary host star/atmospheric molecule combinations.  In each case, the S/N for the detection of the molecule in question is plotted versus the number of transits (solid blue and red lines), and S/N increases as the square root. Our S/N ratios are calculated using multiple Monte-Carlo trials, and the dashed red lines give the $\pm1\sigma$ bounds on our calculations.  The bottom X-axis is only for the bottom panel, and the top X-axis is for the remaining panels. The distance of each system is adopted to be 12 parsecs, and the results from \citet{currie_2023} are from their Figure~13.}
\label{fig: compare}
\end{SCfigure}

Prior to evaluating the effect of star spots, we want to compare our S/N ratios for CC detections to the results from \citet{currie_2023}.  That comparison in four representative cases is shown in Figure~\ref{fig: compare}. This comparison is somewhat of an apples-versus-oranges situation, because the two sets of simulated CC detections have some differences.  Those differences include our different treatment of telluric absorption, our different construction of the template functions, and whether or not to restrict the calculation to specific bands.  In spite of these differences, we find good agreement with \citet{currie_2023}.  We calculate S/N using a Monte Carlo procedure, and that has an intrinsic uncertainty given by the dashed lines ($\pm1\sigma$) on Figure~\ref{fig: compare}.  In the limit of a large number of transits, our agreement with \citet{currie_2023} is often within 1\,$\sigma$, and never worse than 2\,$\sigma$. 

\section{Effect of unocculted star spots}\label{sec: spots}

We now turn to the effect of unocculted star spots, which can be difficult to correct for and which can contaminate and confuse planetary spectra \citep{lim_2023}.  Star spots have two principal effects: 1) they increase the noise in the CC detection of the planet, and 2) in some cases they can contaminate the planetary signal. We emphasize that our calculations are exclusively for {\it unocculted} star spots.  Spots that are crossed during transit can often be seen and corrected using transit spectroscopy data (e.g., \citealp{fu_2024}).  Unocculted spots are difficult to correct for because they can't be measured directly in the data during transit.  Instead, they affect transit spectroscopy via the ``transit light source" (TLS) effect \citep{rackham_2018}, wherein the chord across the stellar disk that is crossed by the planet is not representative of the integrated disk of the star.  The TLS effect enters transit spectroscopy when normalizing the in-transit minus out-of-transit flux, as we now explain. 

If the flux emitted by the star spots does not vary significantly on the time scale of the transit, then the unocculted star spots are not present in the numerator of Eqs.(3) \& (4). ($F_{sp}$ cancels when subtracting $F_{out}$ from $F_{in}$).  The constancy of large spots on the time scale of the transit is supported by studies showing that spot lifetimes increase with the size of the spots \citep{bradshaw_2014}.  Also, \citet{rathcke_2024} successfully used back-to-back transits in the TRAPPIST-1 system to correct for stellar activity; the success of that method implies that the activity on M-dwarf stars like TRAPPIST-1 does not vary strongly on the time scale of a transit. We conclude that the absence of unocculted spots in the numerator of Eqs.(3) \& (4) is a valid approximation.

Unocculted spots only contaminate the transit spectrum when the difference between $F_{in}$ and $F_{out}$ is normalized by dividing by the denominator in Eqs.(3) \& (4).  That normalization is necessary in order to convert the decrease in flux during transit to the fraction of the stellar flux blocked by the planet, and thus measure the apparent radius of the planet as a function of wavelength.  If we knew the flux {\it of the spot-free star} as a function of wavelength to high accuracy at the time of the transit via some other method, then an alternative normalization not containing star spots could be used.  We discuss that possibility in \S\,\ref{sec: alternate}, but we here focus on the effect of the unocculted star spots using Eq.(4).

We calculate the effect of star spots using PHOENIX models for the spots, and a temperature difference and covering fraction in line with the range of values estimated in the literature \citep{giampapa_1985, rackham_2018, bicz_2022, mori_2024, waalkes_2024}.  

\subsection{Increase in noise due to star spots}\label{sec: spot_noise}

\begin{SCfigure}
\centering
\includegraphics[width=3.5in]{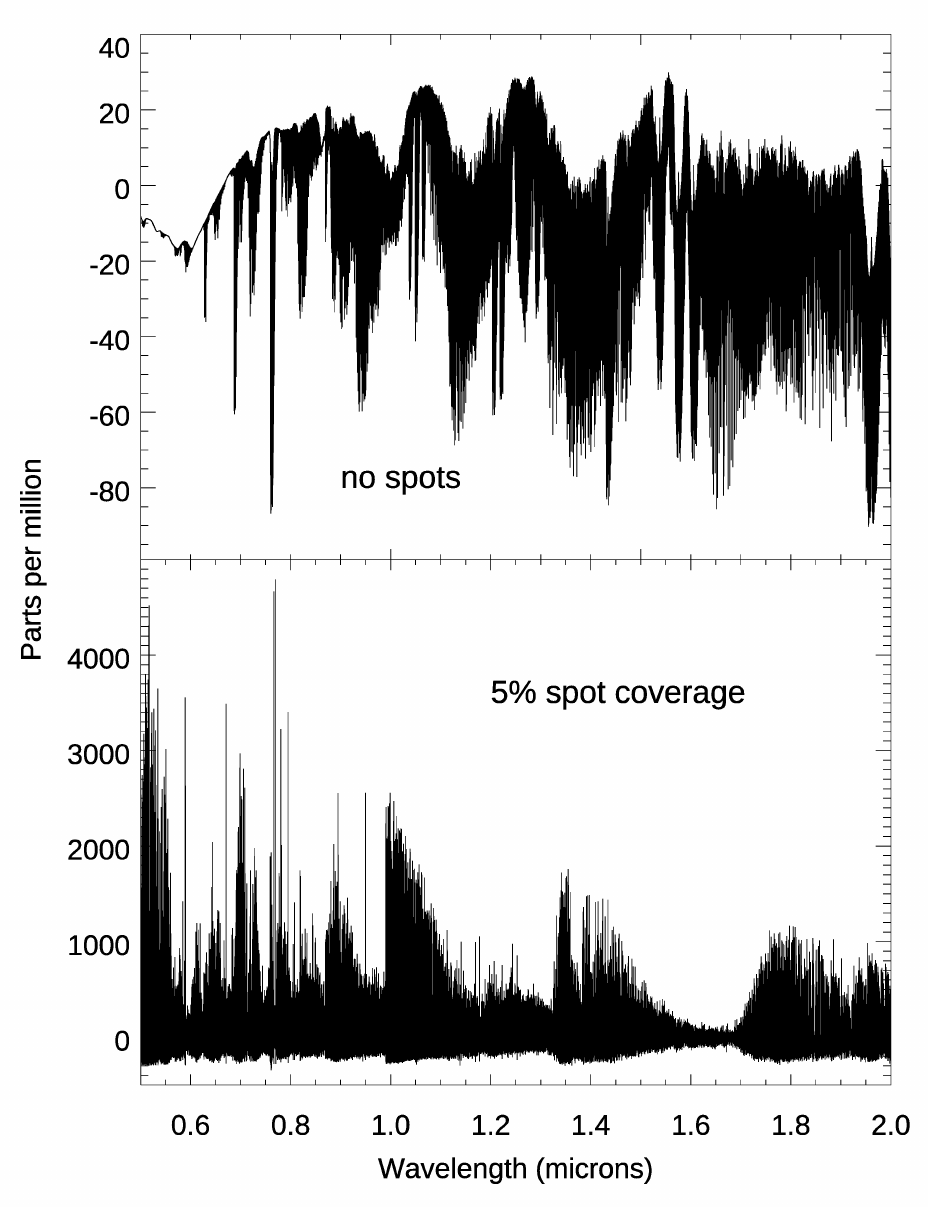}
\caption{Effects of star spots on $S_{\lambda}$ for the case of a temperate planet orbiting an M8V star 12 parsecs from Earth, with 5\% area coverage of an unocculted star spot having $\Delta{T} = -800$\,K.  The upper panel shows the median-subtracted unspotted case, and the effect of spots is shown in the lower panel, where the Y-axis is the median-subtracted $S_{\lambda}$ as per Eq.(4). The effect on the CC function will be smaller than this effect on $S_{\lambda}$ because the star spot spectrum is not highly correlated with the molecular templates.}
\label{fig: spot_effect}
\end{SCfigure}

If no unocculted spots are present, the stellar spectrum cancels in Eq.(4), and only observational photon noise mixes with the exoplanet's transmission spectrum.  In our simulations, we can turn off the photon noise to show how star spots cause additional noise when they don't cancel in Eq.(4), i.e. when $F_{sp}^{\lambda}$ is not zero. Even prior to the CC calculation, the additional noise can be seen in $S_{\lambda}$, shown in Figure~\ref{fig: spot_effect} for the case of an M8V system.  The upper panel corresponds to the unspotted case, and the lower panel shows $S_{\lambda}$ in Eq.(4) for a 5\% spot covering.  

The spectrum of the unocculted star spot can interfere with the detection of molecules in the exoplanetary atmosphere, especially since even modest spot coverage can produce contamination that is orders of magnitude higher than the planetary molecular signal. In the unspotted case, Figure~\ref{fig: spot_effect}, $S_{\lambda}$ varies by tens of ppm, due exclusively to transmission by a mixture of molecules in the exoplanetary atmosphere. However 5\% coverage by unocculted star spots (having $\Delta{T} = -800$\,K) in the lower panel introduces fluctuations that are larger by more than an order of magnitude.  Strictly speaking, those fluctuations are not noise, they are signal from the spectrum of the star spot.  However, they function as noise for the CC calculation.  

\begin{SCfigure}
\centering
\includegraphics[width=3.5in]{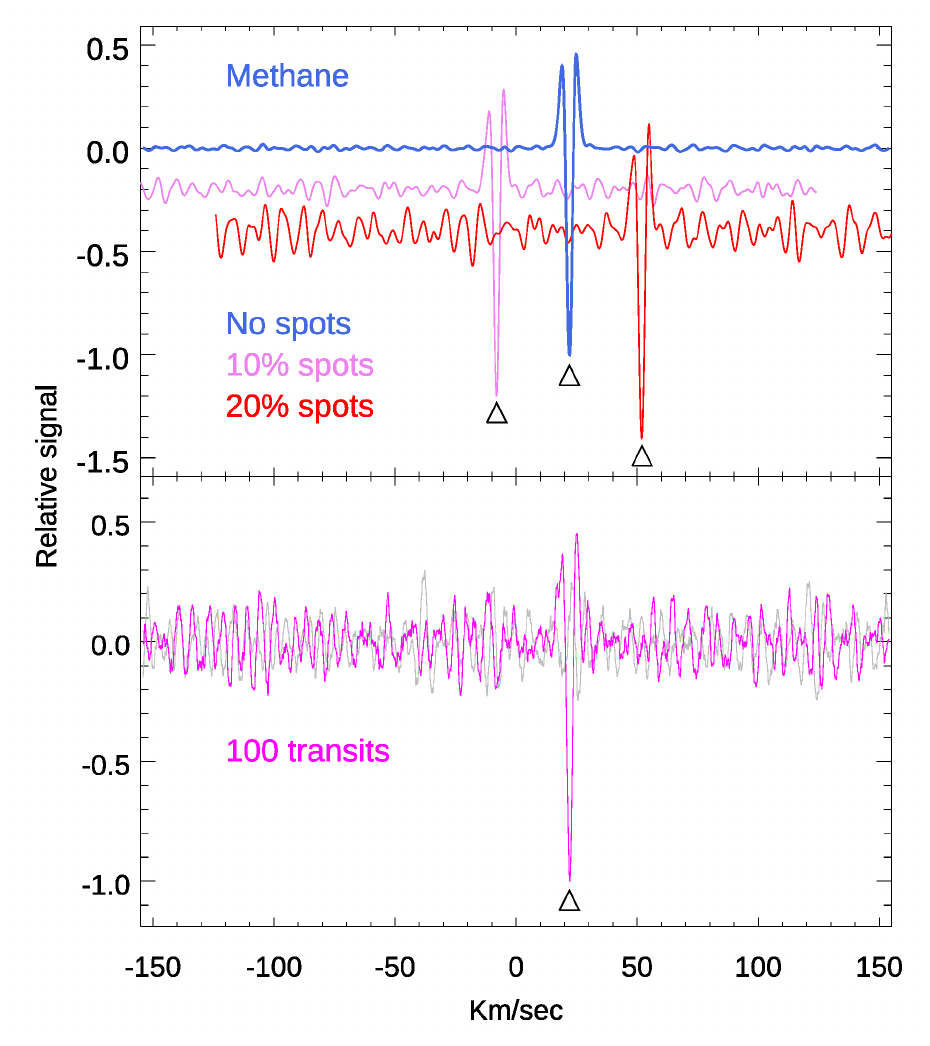}
\caption{Simulated cross-correlation (CC) detections of methane in a temperate planet orbiting an M4V star 12 parsecs from Earth, using the 39-meter E-ELT.  The upper panel shows the CC functions with observational photon noise turned off, and various coverage of unocculted spots having $\Delta{T}=-400$\,K. The CC functions are displaced in velocity for clarity (avoiding overlap), and the triangles mark their centers. In the absence of star spots and photon noise (blue curve) the S/N for the methane detection (=145) is limited by overlap with other molecules in the exoplanetary atmosphere.  When limited by unocculted star spots, the S/N is 36 and 18 for 10\% and 20\% spot coverage, respectively.   The lower panel adds photon noise to the unspotted case, using the E-ELT ($978\,m^2$ collecting area with 10\% efficiency), and 100 transits. The S/N for that detection is 12.}
\label{fig: methane_fig}
\end{SCfigure}

We have propagated the fluctuations in $S_{\lambda}$ that are caused by star spots to the noise that they produce in the CCs.  An example is shown in Figure~\ref{fig: methane_fig} for the detection of methane in a temperate rocky planet orbiting an M4V star.  The star spot is 400\,K cooler than the photosphere, and we describe the area coverage of the spot as a fraction of the stellar disk when projecting to the plane of the sky.  A spot coverage of 10\% produces noise in the CC at a level less than the photon noise in 100 transits (lower panel).  Only for spot coverage of 20\% does the spot-induced noise come close to the amplitude of the photon noise.  Those fractional spot coverages are large: a 10\% area coverage corresponds to a single star spot with an (assumed circular) projected radius more than 30\% of the stellar radius.

\begin{SCfigure}
\centering
\includegraphics[width=3.5in]{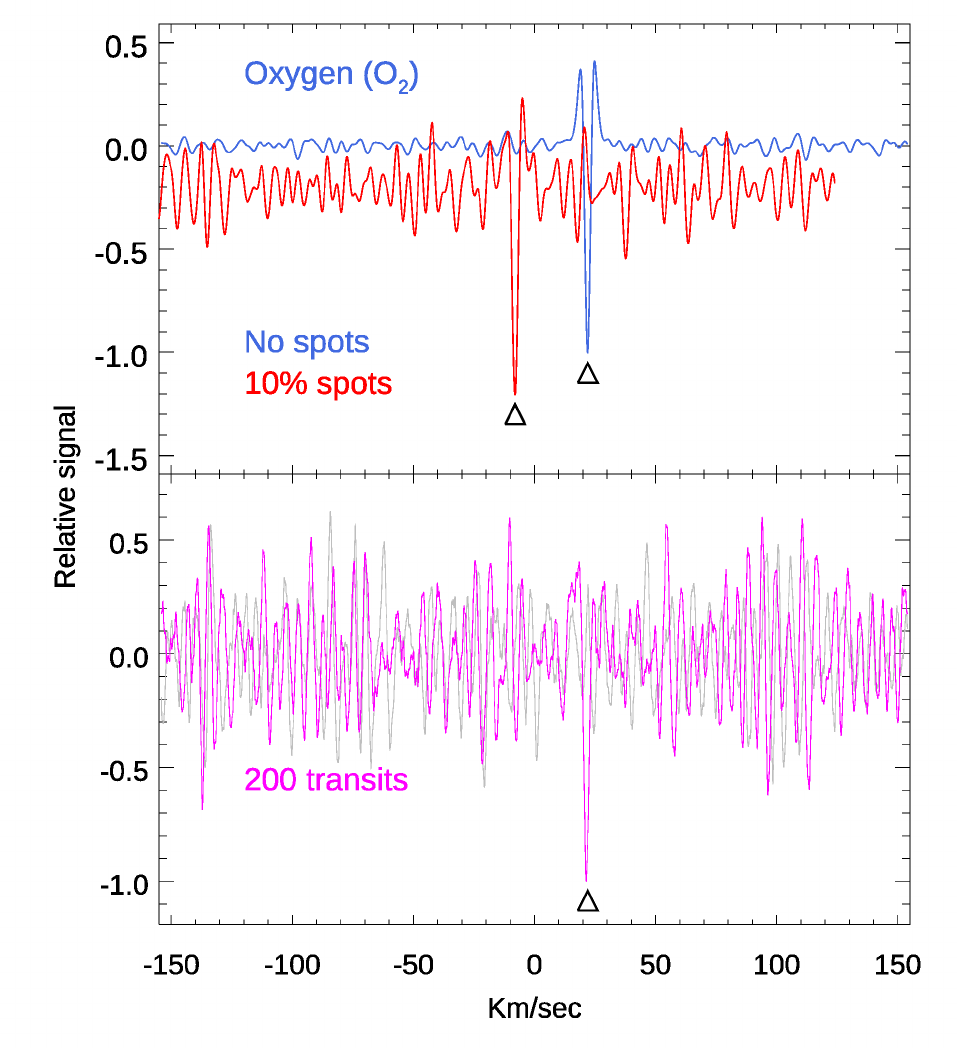}
\caption{Simulated cross-correlation (CC) detections of molecular oxygen in a temperate rocky planet orbiting an M4V star, 12 parsecs from Earth, using the 39-meter E-ELT.  This is the same as Figure~\ref{fig: methane_fig}, but for the case of molecular oxygen.  In the absence of unocculted star spots (blue curve) and with photon noise turned off, the S/N for the detection is 43, limited by overlap between the template and other molecules present in the exoplanetary atmosphere.  With 10\% unocculted spot area, the S/N is 9, and when considering only photon noise for 200 transits (lower panel) the S/N for the detection is 4.}
\label{fig: oxygen_fig}
\end{SCfigure}

Figure~\ref{fig: oxygen_fig} shows the same star/planet case, but for molecular oxygen.  Molecular oxygen is more difficult to detect than methane because it requires more transits \citep{lopez-morales_2019, currie_2023}, and high spectral resolution is crucial \citep{lopez-morales_2019,hardegree_2023}. Nevertheless, \citet{snellen_2013} projected detection of molecular oxygen in a few dozen transits, in a favorable case.  We find that star spots will not prevent the detection: 10\% coverage by star spots produces noise less than the photon noise averaged over 200 transits.  We point out that the noise produced by star spots will probably not average down in proportion to the square-root of the number of transits, as would pure photon noise.  The star spot noise is not stochastic, it's due to spectral structure that will correlate to a large degree in different star spots. Nevertheless, it doesn't have to average down if, as we calculate, it remains below the levels of photon noise that permit the detection of carbon dioxide, methane, and molecular oxygen.

\begin{figure}
\centering
\includegraphics[width=4in]{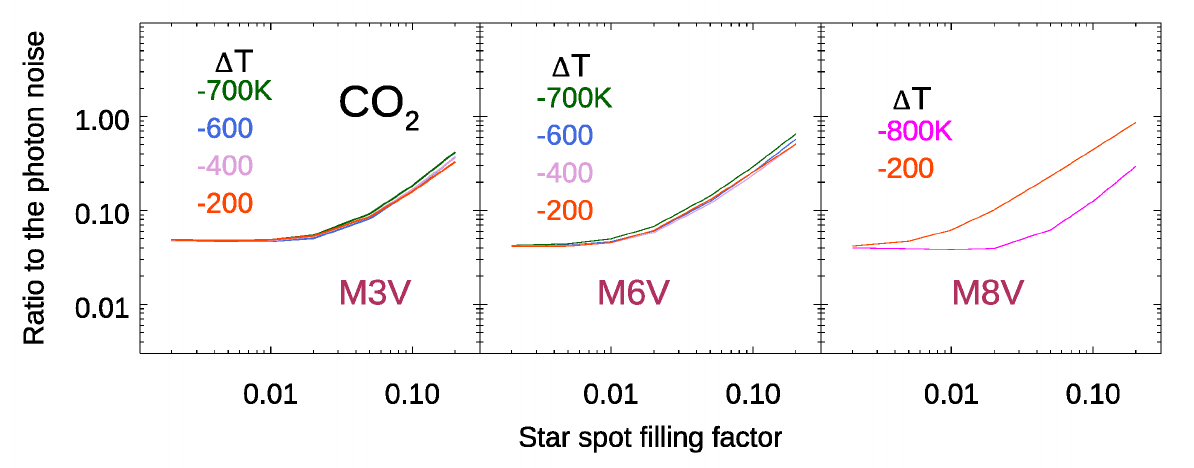}
\caption{Ratio of noise from unocculted star spots to the photon noise when averaged over 100 transits. These are for detection of carbon dioxide in a temperate planet transiting M3V, M6V, and M8V stars using the E-ELT, as a function of the filling fraction of the spots.  The different curves are for different temperature differences between the photosphere and star spots.}
\label{fig: cnoise_CO2}
\includegraphics[width=4in]{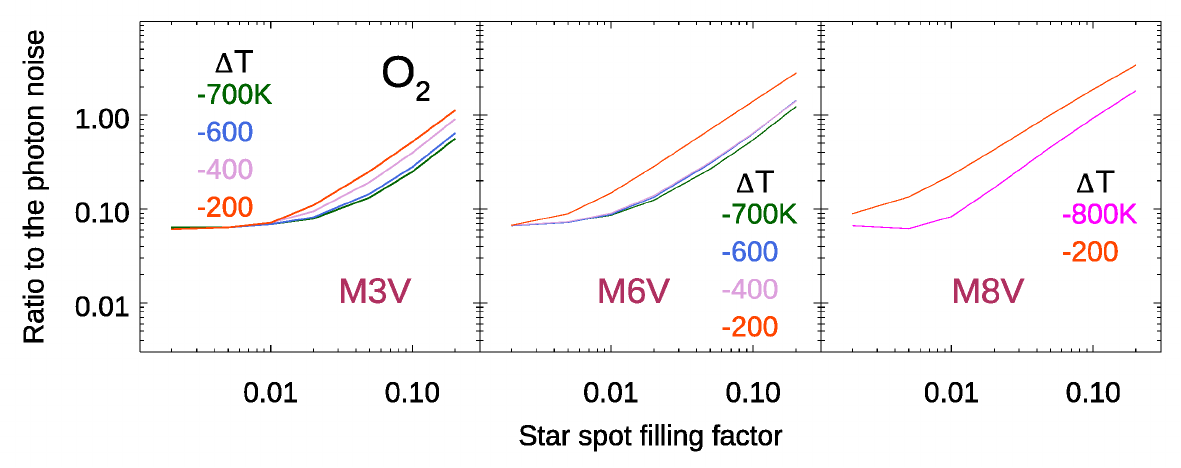}
\caption{The same calculations as in Figure~\ref{fig: cnoise_CO2}, but for the case of molecular oxygen.}
\label{fig: cnoise_oxygen}
\includegraphics[width=4in]{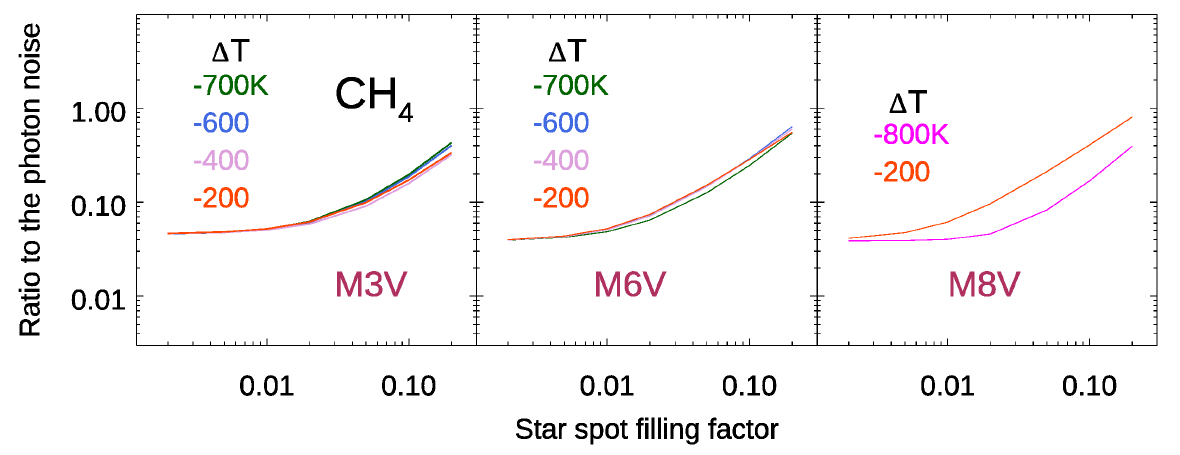}
\caption{The same calculations as in Figure~\ref{fig: cnoise_CO2}, but for the case of methane.}
\label{fig: cnoise_CH4}
\includegraphics[width=4in]{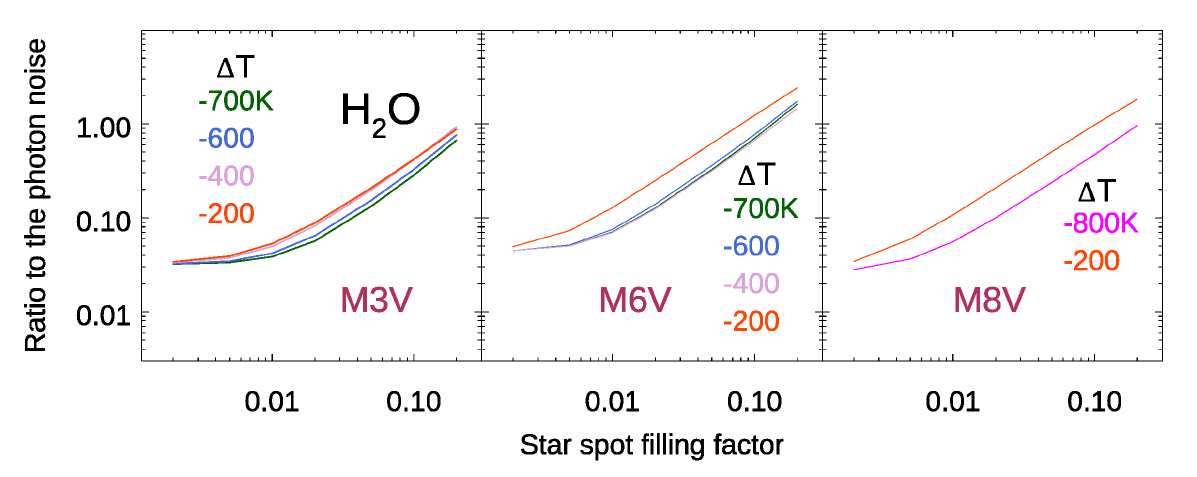}
\caption{The same calculations as in Figure~\ref{fig: cnoise_CO2}, but for the case of water vapor.}
\label{fig: cnoise_H2O}
\end{figure}

We have calculated the ratio of star spot noise to photon noise for all four molecules we study, for M3V, M6V, and M8V spectral types, and as a function of the star spot filling fraction.  Those results are shown for the average of a 100-transit program in Figures~\ref{fig: cnoise_CO2}, \ref{fig: cnoise_oxygen}, \ref{fig: cnoise_CH4} and \ref{fig: cnoise_H2O}. We used high resolution PHOENIX models from \citet{husser_2013} to represent the star spots, but those models are not available at temperatures below 2300\,K.  Consequently, for the M8V case we calculated a model at 1700\,K, using the methods described by \citet{peacock_2019a, peacock_2019b, peacock_2020}.  We thereby bracket the star spots for the M8V case using two models (1700 and 2300\,K).

Some general trends are evident on Figures~\ref{fig: cnoise_CO2} to \ref{fig: cnoise_H2O}.   First, the noise produced by star spots increases with increasing filling fraction, and the filling fraction overall has a larger effect than does the star spot temperature.  Interestingly, spots having the smallest temperature difference with the photosphere ($\Delta{T}=-200$\,K) often produce the largest noise, especially for detection of molecular oxygen.  That occurs because the hottest star spots emit the greatest flux, and that high flux more than compensates for their simpler spectral structure as compared to cooler spots.  To view that another way, consider what must happen in the limit where the spot temperature decreases greatly, i.e., a very cold spot. In that limit, the spot spectral structure becomes undetectable, and the spot only blocks photospheric flux but adds no detectable photons.  A very cold spot thereby produces only an offset in the transit depth, but adds no noise to the CC.  That effect would be most pronounced for molecular oxygen, whose relatively short wavelengths of absorption (0.76 and 1.27\,$\mu$m) are on the steep side of the Planck function at these temperatures.  Consequently, it is reasonable that warmer spots produce more CC noise for oxygen than do cooler spots.    

For all four molecules, star spot noise becomes significant compared to the photon noise averaged over 100 transits, but the star spot noise does not dominate sufficiently to prevent detections of the molecules.  For example, for methane observed in an M3V system (Figure~\ref{fig: cnoise_CH4}), 10\% spot coverage - a large amount - produces noise in the CC equal to about 0.2 of the photon noise.  Hence the star spot noise would degrade that detection by a factor of $\frac{1}{\sqrt(1^2+0.2^2)} = 0.98$, only a 2\% decrease in S/N.

\subsection{Contamination of water vapor by star spots}\label{sec: contamination}

\begin{SCfigure}
\centering
\includegraphics[width=4in]{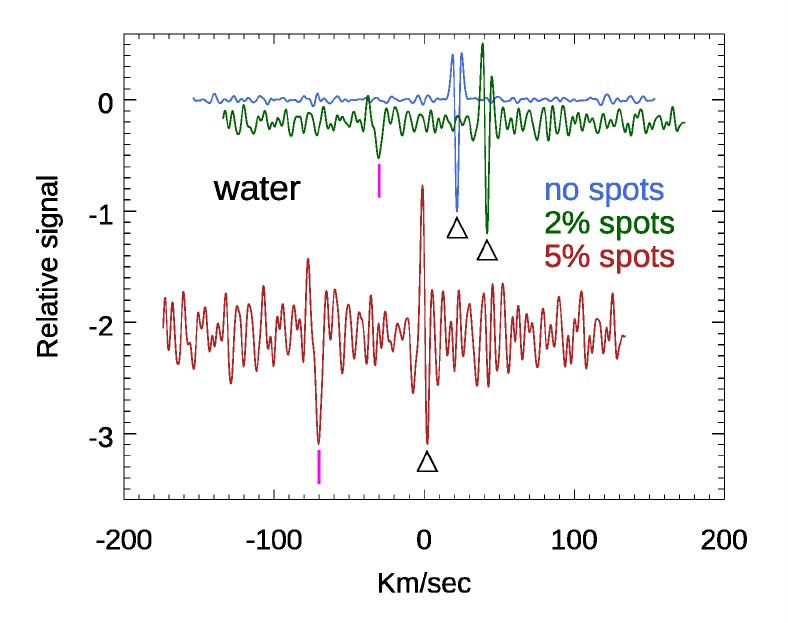}
\caption{Effect of star spot water vapor on cross-correlation (CC) detections of exoplanetary water.  Three CC functions are shown, for no spots, and 2\% and 5\% spot filling fractions.  The CC functions are displaced in velocity (and also in the Y-axis) to illustrate them without overlap, and the star spot velocity is set at -70 km/sec relative to the planet, so that their CC peaks do not overlap. (In the case of most real M-dwarf stars, the CC peaks from the planet and star spot would be closely coincident).  Triangles mark the signal from water in the exoplanet, and the magenta lines mark the signal from water in the star spot. This is for the M6V case, with star spots 600\,K cooler than the photosphere. \label{fig: waterseq} }
\end{SCfigure}

At the temperatures of star spots, only the simplest molecules are believed to be present in detectable amounts \citep{grevesse_1973}.  Of the four molecules we study, only water vapor will be detectable in star spots.  It is well known to be present in sun spots \citep{wallace_1995}, and in the photospheres of cool stars \citep{jones_1995}. Even assuming that telluric water absorption can be efficiently canceled (as our simulations adopt), star spot water absorption will overlap and contaminate the exoplanetary transit signal.  The rotational periods of most M-dwarf stars fall in the range from 10- to 30 days \citep{newton_2016, popinchalk_2021}, and consequently the radial velocity of the spots will be close to that of the stellar center of mass. To investigate the contamination by star spot water vapor, we adopt an (unphysical) radial velocity of -70 km/sec for the spot relative to the star. That cleanly separates the star spot water from the exoplanetary water, and allows us to evaluate them separately, as illustrated in Figure~\ref{fig: waterseq}. 

Figure~\ref{fig: waterseq} shows CC functions for water with the photon noise turned off, and with two star spot filling fractions. Water in the spot produces a CC peak (marked by magenta lines), that  becomes equal to the exoplanetary water signal for 5\% filling fraction.  The star spot appears in the CCs even though our template function is characteristic of 296\,K, whereas star spot water is much hotter. 

Star spots not only produce their own peaks in the CC functions, but they also increase the noise level for water vapor more than for other molecules.  Not only is that visible in Figure~\ref{fig: cnoise_H2O}, but it is especially apparent in Figure~\ref{fig: waterseq}.  That Figure shows that a 5\% filling fraction of star spots increases the noise to the point where the exoplanetary water is only marginally detectable. (The exoplanet water marked by the triangle below the red CC is only marginally beyond the noise envelope).  Moreover, that detection would be compromised by blending of the exoplanetary water with the star spot water, because in real detections the star spot would not be Doppler-shifted to the degree we illustrated in Figure~\ref{fig: waterseq}. Specifically, the rotational velocity of an M-dwarf star of this radius is typically no more than 2 km/sec \citep{newton_2016}, and the water CC signal from the spot will overlap the exoplanet water signal. However, we have developed a technique to minimize the effects of star spots, and maximize the detectability of water vapor, as well as other molecules.

\section{Alternate normalization}\label{sec: alternate}

The effects of unocculted star spots on transit spectroscopy do not appear directly in the difference between in-transit and out-of-transit flux, i.e. they are not in the numerator of Eq.(4).  Instead, they appear when the flux difference is normalized to be a fraction of the star's total flux, i.e. the unocculted spots are introduced by the denominator of Eq.(4).  We propose an alternate method of normalization that minimizes the impact of unocculted spots.  Although we here describe this method as applied to CC detections, it is equally applicable to other forms of transit spectroscopy.  

Alternate normalization derives a proxy spectrum for the star without spots, and uses that spectrum as the denominator of Eq.(4).  We illustrate the method using an M6V star, with a variable fraction of star spots, and we adopt a temperature of the spots that is 600\,K cooler than the quiet photosphere.  Adopting that the spot filling fraction varies with time, our simulation obtains high resolution spectra of the star on 10 occasions out of transit, each observation achieving a S/N of 300.  (That S/N is high, but reasonable for the ELTs.)  We also adopt that the star's V-K photometric color is measured on those 10 occasions, to a precision of 0.003 magnitudes (also reasonable for current ground-based capabilities).  We choose the spot filling fraction to vary randomly using a uniform distribution between 0.0 and 0.15, which spans the range commonly observed for M-dwarfs (e.g., \citealp{rackham_2018}). Using PHOENIX models for the quiet photosphere and spots, we calculate the V-K colors of the star on the 10 observed occasions.  (We choose 10 observed samples as a compromise between the precision of the final result, and what is achievable in a reasonable observing program.).   

\begin{SCfigure}
\centering
\includegraphics[width=3in]{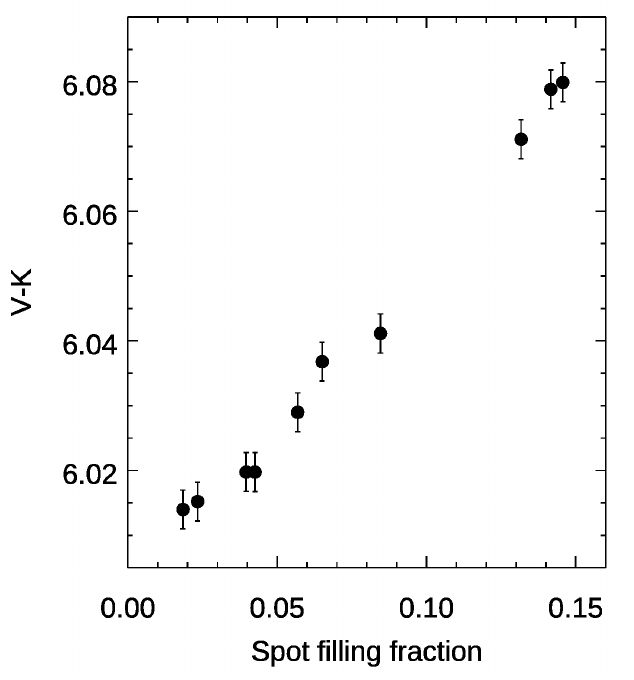}
\caption{V-K photometric color (in magnitudes) of an M6V star versus the filling fraction of star spots, adopting that the spots are all 600\,K cooler than the quiet photosphere, and varying their filling fraction randomly from 0.0 to 0.15. \label{fig: VK} }
\end{SCfigure}

\begin{figure}
\centering
\includegraphics[width=5in]{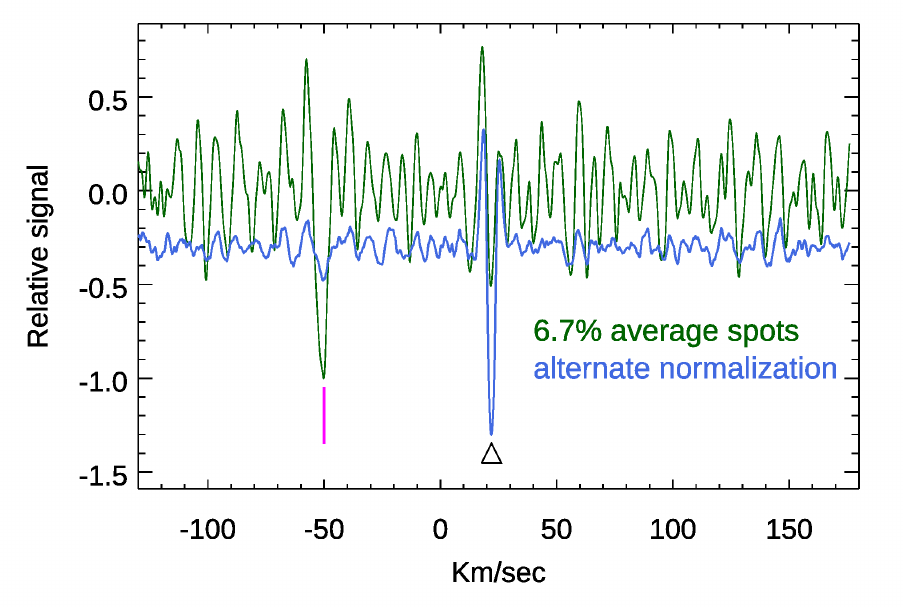}
\caption{Effect of alternate normalization to reduce the noise and contamination from star spots in a CC detection of water vapor for an M6V host star.  In this example, photon noise has been turned off, and the noise in the CC is due to the star.  The 6.7\% average degree of spottedness produces a large amount of noise that prevents detection of the dip in the CC due to the planet, and the spots also produce their own dip, that we moved to -50 km/sec (marked by magenta line) to avoid confusion with the exoplanet dip.  The result from alternate normalization shows a factor of 6 reduction in noise, and contamination of the exoplanetary signal is less by a factor of 9 (at $3\sigma$ significance). \label{fig: alternate} }
\end{figure}

Figure~\ref{fig: VK} shows our calculated V-K color versus spot filling factor.  With increasing filling factor, the spots make the star as a whole effectively cooler, and it becomes redder, i.e. V-K increases.  That agrees with population studies finding that M-dwarfs become redder in V-K with decreasing T$_{eff}$ (e.g., Figure~1 of \citealp{mignon_2023}).  In order to produce a spot-free proxy spectrum, it is important to work at high spectral resolution.  We thereby process the 10 simulated high resolution spectra one wavelength at a time.  At each wavelength, we regress the flux of the spotted stars versus their V-K color, using simple linear least-squares solutions.  Extrapolating those linear solutions to the V-K color of an unspotted star, defines the flux of the proxy spectrum at each wavelength.  Figure~\ref{fig: alternate} shows - for at least this simple case - that using the proxy spectrum for an alternate normalization reduces the noise from the unocculted spots by a factor of 6, and reduces their contamination of the exoplanetary water vapor CC signal by a factor of 9 (at $3\sigma$ significance).

In our simulations, we know the V-K color of an unspotted star from the PHOENIX models, so we can extrapolate the spotted spectra to that color. Real observations would invoke population studies to find the unspotted V-K, e.g., by observing how V-K varies with activity indicators such as the strength of emission in the Ca K-line \citep{meunier_2024}. Also, our linear regressions use fluxes from the PHOENIX models, with random noise added, but without systematic errors.  Real observed spectra could contain systematic errors in observed fluxes due to uncalibrated instrumental effects.  In that case, the method would use fluxes relative to a continuum, and relative fluxes of that type would be insensitive to fluctuations in instrumental sensitivity. Robust continuum-placement methods have been developed, such as the RASSINE method \citep{cretigner_2020}.  Thus, we deem our method of alternate normalization to be viable, but some additional caveats are discussed in \S\,\ref{sec: caveats}.  Moreover, although we here illustrate the method using PHOENIX models, in a real application it would use only observed spectra and would therefore be independent of uncertainties in the models \citep{rackham_2024}.  Finally, we acknowledge that alternate normalization is similar to the method described by \citet{zellem_2017}, and used to predict the impact of stellar variability on JWST transit observations, but is different in the details of how it is implemented.

\section{Challenges to implementation}\label{sec: caveats}

Applications of this work to real planetary systems must overcome several challenges, and we here briefly list several of those challenging aspects. \S\ref{sec: comprehensive} further discusses a comprehensive approach to successfully deal with some of these challenges.
\begin{itemize}
\item {\bf Planets can be cloudier than we expect.} We used the exoplanetary transmission models from \citet{currie_2023}, and their pre-industrial Earth model transmission spectra contain a mixture of cirrus and stratocumulus clouds.  Indeed, clouds have proven to be of major importance and a limiting factor in transit spectroscopy of many planets from gas giants to rocky worlds \citep{wakeford_2019a, komacek_2020, lustig-yaeger_2023}.  It is possible that rocky exoplanets observed with the ELTs may have more clouds than in the models, thus reducing the sensitivity to molecular detections via transit spectroscopy. However, our goal in this work is to define the effect of unocculted star spots, and not to explore additional cloud effects, so we have focused on the nominal models used by \citet{currie_2023}.

\item {\bf Occulted star spots can still be problematic.}  Although this paper is focused on the case of UNocculted star spots, smaller occulted star spots can still be problematic.  Large spots that are crossed during transit can often be corrected during data analysis \citep{fu_2024, fournier_2025}, but small spots that are crossed may be individually undetectable.  In aggregate, small spots crossed during transit could interfere with exoplanetary molecular detections, especially when many transits are averaged. A potential solution to that contamination is to obtain direct measurements of the spectral properties of star spots and faculae, as we discuss in \S\ref{sec: deriving}. Those measurements would allow for a more accurate inclusion/correction of star spot and facular contamination in transit spectroscopy retrievals.

\item {\bf Star spots might not vary sufficiently.}  Alternate normalization relies on measuring the variability of the star spot covering fraction, e.g., as the star rotates.  However, the method would fail if star spots were uniformly distributed in longitude, not transited by the planet, and long-lived.  In that case there could be insufficient temporal variation in the star's V-K color to enable an extrapolation to the unspotted case.  However, we deem this to be an unlikely situation, and we note that M-dwarfs commonly exhibit significant photometric variations due to their star spot distributions and rotation \citep{newton_2016}, and also variability on long time scales \citep{weis_1994}. We expect that those variations could in most cases be successfully extrapolated to the unspotted case, as we discuss in \S\ref{sec: never}.

\item {\bf Unocculted star spots can be elusive.}  Alternate normalization extrapolates variations in spottedness to the unspotted case, and we have demonstrated (\S\,\ref{sec: alternate}) that it can be successful in principle for a simple case.  In actual practice, it is equivalent to measuring the spectra of the unocculted star spots.  However, measuring high resolution spectra of unocculted star spots has not (so far) been successful in actual practice: spot temperatures and filling factors have been measured (e.g., \citealp{cao_2022}), and spectral energy distributions \citep{fournier_2025}, but not high resolution spectra of the spots. For example, \citet{morris_2020} did not detect unocculted spots using a CC technique, even on the very spotted star HAT-P-11.  Those authors regarded that as good news, because it suggested minimal contamination of planetary emission spectroscopy via cross-correlation, but we regard it as bad news, because spot spectra remain unmeasured.   We have no doubt that unocculted spots can contaminate transit spectroscopy of exoplanets \citep{rackham_2023, fournier-tondreau_2024, fournier_2025}.  If our alternate normalization is to be successful to minimize that contamination in real observations, then we deem it necessary to demonstrate a detection of the spectra of the unocculted spots at high spectral resolution, and we discuss how to do that in \S\ref{sec: deriving}.

\item {\bf Faculae may be dominant.}  Star spots {\it per se} typically cover less of the star than do faculae (e.g., \citealp{rackham_2018}).  Moreover, because faculae are brighter than the quiet photosphere \citep{delaroche_2024} they could have a dominant effect on CC detections.  One indication of that is that we sometimes find that the hottest star spots produce the greatest noise compared to cooler spots (Figure~\ref{fig: cnoise_oxygen}).  We briefly explored the effect of adding faculae to our alternate normalization method, and we find that adequately correcting for a {\it mixture} of faculae and spots would require a much higher S/N (on Figure~\ref{fig: VK}) than is likely to be possible. In addition to V-K color, proxy data that are specifically sensitive to the coverage by faculae may be needed. The magnetic structures that define faculae expand with height, and become bright plage regions in the chromosphere.  Bright chromospheric plage are thereby manifestations of the same physical process that defines faculae \citep{chatzistergos_2022}.  Spectral indicators that are sensitive to chromospheric emission \citep{sowmya_2024} are potentially valuable data that could be used to improve our alternate normalization method by accounting for faculae (discussed further in \S\ref{sec: alt_diagnostics}).  In this regard, we are encouraged by recent work showing that magnetic activity on the Sun is ``compact and coherent in the spectral domain" \citep{zhao_2024}.  Consequently, only a few spectral indicators may suffice to derive spot-free proxy spectra.  
\end{itemize}

\section{A Comprehensive Approach}\label{sec: comprehensive}
We here discuss possible solutions to many of the challenges highlighted in \S\ref{sec: caveats}.

\subsection{Alternate diagnostics}\label{sec: alt_diagnostics}

Although our simple extrapolation to the spot-free case (\S\ref{sec: alternate}) used the stellar V-K color, other photometric colors could be explored.  \citet{amado_1999} found that the I-K color was effective to measure the average stellar surface brightness, and it would thereby be sensitive to spots and faculae.  Multi-band photometry \citep{rockenfeller_2006} may also be effective in constructing an optimal photometric indicator of M-dwarf activity, especially considering that spots and faculae can have opposite effects on stellar brightness.  Work on L-dwarfs \citep{oliveros-gomez_2024} has shown that customized bandpasses can optimize the sensitivity to variability. In addition to photometry, spectroscopy of the Balmer lines \citep{garcia_soto_2025} or other strong lines such as CaII $\lambda$\,8542 \citep{ivanova_2004} could be used as a proxy for magnetic activity. Finally, spectroscopic indicators such as line equivalent widths that are affected by Zeeman broadening could potentially be used in lieu of photometry \citep{muirhead_2020}.

\subsection{Limitations of the stellar models}\label{sec: model_lim}
Alternate normalization requires constructing a proxy spectrum for the M-dwarf stellar host that is free of the spectral signatures of magnetic activity.  However, M-dwarf stars have complex emergent spectra, and it can be challenging for models to reproduce their observed line spectra \citep{rice_2009} at high resolution to within the observational errors.  Fortunately, recent advances in molecular line data \citep{tennyson_2024}, as well as model atmospheric structure \citep{iyer_2023} and data-driven techniques \citep{ness_2015}, are improving this situation.  For example, see Figure~2 of \citet{may_2023} for good model replication of an M-dwarf spectrum.  Another example is \citet{behmard_2025}, who used a data-driven predictive model of M-dwarf stellar fluxes to derive elemental abundances with uncertainties better than 0.03 dex. These advances are encouraging for the achievement of alternate normalization in actual practice, leveraging both improved understanding of M-dwarf atmospheric structure as well as data-driven extrapolation methods that are more powerful than the simple technique we discussed in \S\ref{sec: alternate}.

\subsection{Complex spot configurations}\label{sec: complex}

In \S\ref{sec: alternate}, we demonstrated an alternate normalization using a single homogeneous star spot temperature and filling factor.  In actual practice, sunspot \citep{cortie_1901} and star spot configurations \citep{afram_2019} can be complex.  In that case, the spatial variations in star spot magnetic field strength and temperature could potentially frustrate alternate normalization if only simple proxies such as photometric colors are used.  Direct magnetic indicators such as Zeeman broadening \citep{shulyak_2019, muirhead_2020} may help to deal successfully with magnetic complexity.

\subsection{Never unspotted?}\label{sec: never}

It is possible than a given M-dwarf star hosting transiting planets may never be unspotted.  If so, how can we extrapolate its measurements to a spot-free case that never occurs?  We point out that extrapolations have successful precedent in astronomy.  For example, in ground-based observational photometry, observers routinely derive magnitudes and colors of stars by extrapolating measurements to zero airmass.  Although ground-based astronomers never observe a star at zero airmass, the extrapolations are successful because magnitudes have a known functional relation to airmass, and airmass can be determined precisely and unambiguously.  In the case of spotted M-dwarf stars, the physics is different but an extrapolation may be possible using Zeeman effects.  Zeeman broadening is a known function of magnetic field strength, and field strength drives the temperature fluctuations in active regions (both for spots and faculae).  Using the broadening of Zeeman-sensitive lines (\S\ref{sec: alt_diagnostics} \& \S\ref{sec: complex}) formed at different stellar atmospheric depths can potentially enable a successful extrapolation to the unspotted case. 

\subsection{Deriving spot and facular properties}\label{sec: deriving}

Alternate normalization relies on extrapolating stellar spectra to the magnetic-free state (\S\ref{sec: alternate}).  To the extent that the extrapolation could not be made perfectly, information on the nature of the star spot spectra would be valuable as strong priors for modeling star spot contamination of exoplanet transit spectra.  Ratios of high resolution spectra taken under different degrees of spottedness (rotational phases) can probe the structure of the star spot and facular spectra.  It may thereby be possible to derive strong constraints on the spectra of star spots and faculae using cross-correlation techniques. In some cases, star spot and facular signatures may not vary strongly with rotational phase as we have modeled below.  In those cases, spots and faculae on other stars of similar spectral type with strong rotational variations could be probed.  At a minimum, that information could clarify how real magnetic structures differ from non-magnetic models (e.g., PHOENIX).

We have simulated the process that we envision, using PHOENIX models.  Non-magnetic models such as PHOENIX can only represent temperature fluctuations, i.e. colder than the quiet photosphere for star spots and hotter for faculae, but they have enabled advances in the solar case \citep{fontenla_2011}.  In any case, non-magnetic models are sufficient to illustrate the observational and analysis process that we envision.  The method is quite general and would be able to probe the real spectra of active regions.  Figure~\ref{fig: phoenix_cc} shows the ratio of a spotted M-dwarf at high spectral resolution (R=80,000, using PHOENIX models \citealp{husser_2013}) to the same M-dwarf with the spots rotated out of sight.  We show a small representative portion of the red-optical spectral ratio near the strong resonance line of neutral potassium (KI) at 7667\,Angstroms.  The potassium line being an absorption from the ground state, it is stronger in the star spot than in the quiet photosphere, and it appears as extra absorption in the ratio spectrum.  Similar behavior is seen in other relatively low excitation lines (ThI 1.0eV of excitation, and TiI 2.5 eV).  Whereas lines whose lower states lie at higher excitation (MgI 5.1 eV, FeI 3.0eV, and CrI 6.2eV) are stronger in the quiet photosphere than in the spot, and they appear like emission lines in the ratio.  

Key information on the excitation state of star spots (and also faculae) could be obtained using a cross-correlation technique. We added observational noise{\footnote{based on achieving S/N = 200 using the MAROON-X spectrometer on the Gemini-North telescope, \url{https://maroon-x-etc.gemini.edu/app}} to the spectral ratio in Figure~\ref{fig: phoenix_cc}, and applied a high-pass filter in wavelength to isolate the line structure. The noise is not plotted on the figure, but the ratio of two spectra each having S/N = 200 would produce a standard deviation in the ratio spectrum of $\sigma=0.007$. Using the noiseless modeled spectrum as a template, we calculate the CC function for the spectrum ratio plus noise.  That CC is shown as the inset on Figure~\ref{fig: phoenix_cc}; it has S/N=90 for detection of the total line structure in the ratio spectrum over the full wavelength range (6000 to 9000 Angstroms). That should be an easy detection for a program of high resolution spectroscopy. The individual lines in the spectral ratio on Figure~\ref{fig: phoenix_cc} will be beneath the noise level, but the CC process effectively averages over many lines and can produce a high-S/N detection.  Having S/N=90 for the totality of the spectral structure leaves ample dynamic range to probe subsets of the line structure.  For example, a template constructed using high-excitation lines could probe facular regions, whereas templates focused on low-excitation lines could probe spot umbrae.  Even if an unspotted reference phase is not available, the variations in spectral structure will be informative. Based on current MHD modeling \citep{witzke_2022, smitha_2025} we expect that the variations in spectral structure will deviate significantly from non-magnetic models. High resolution spectroscopy of planet-hosting stars at multiple rotational phases, combined with a CC analysis that is tuned to spectral features of differing excitation states, thereby has great potential to inform us concerning the spectra of active regions on M-dwarfs. 

\begin{figure}
\centering
\includegraphics[width=4in]{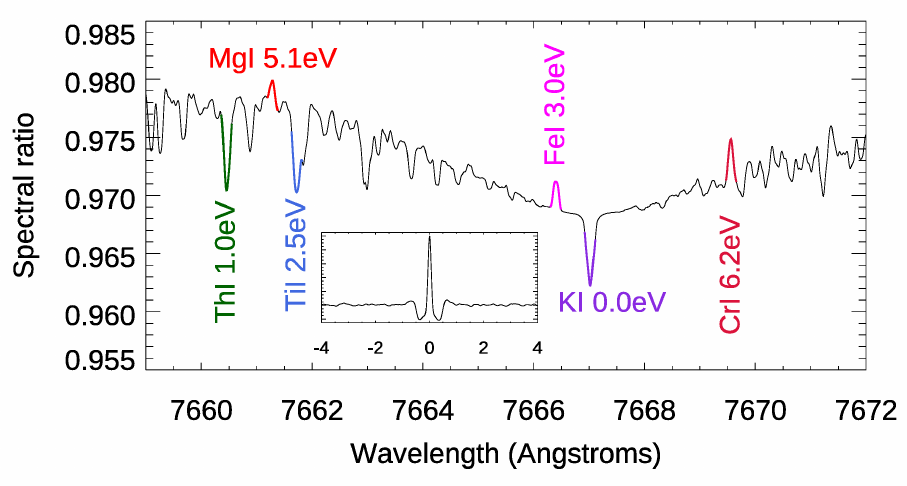}
\caption{Ratio of the spotted spectrum of an M2V star to the same star without spots, using PHOENIX models \citep{husser_2013}. (Only a small portion of the spectrum near a resonance line of neutral potassium is shown.) The spots have 5\% filling factor and ${\Delta}T=-300$\,K.  Colors indicate several atomic lines of different lower state excitation in electron volts (eV). The lowest excitation lines appear as extra absorption in the spectral ratio, whereas the higher excitation lines appear like emission (because they are weaker in the spot than in the quiet photosphere).  The inset plot shows a cross-correlation detection of the line structure in this spectral ratio, using realistic observational noise (noise is not plotted, see text).  Wavelengths are in vacuum. }
\label{fig: phoenix_cc}
\end{figure}

\section{Summary}\label{sec: summary}

This paper is a follow-up to \citet{currie_2023} who simulated the detectability of molecules in the atmospheres of rocky temperate planets transiting M-dwarf stars, from M2V to M8V spectral types, using the ELTs.  Like \citet{currie_2023}, we use a cross correlation (CC) technique to simulate the detections, and we focus on four molecules: carbon dioxide, methane, molecular oxygen, and water vapor.  We use the same planetary atmospheric models as did \citet{currie_2023}, but our calculations differ in some aspects from their work. Nevertheless, we broadly confirm their S/N projections (\S\,\ref{sec: comparison}).  We expand on their work by including the effects of unocculted star spots.  Starting with simple analytic relations (\S\,\ref{sec: formulae}), we point out (\S\,\ref{sec: spots}) that unocculted star spots cancel when subtracting fluxes measured in-transit from out-of-transit fluxes, assuming that the spots do not vary significantly on the time scale of the transit.  The effects of unocculted star spots are introduced when the in-transit minus out-of-transit fluxes are divided by the flux of the star in order to define the fraction of stellar flux blocked by the planet as a function of wavelength.  That normalization allows unocculted spots to introduce noise into the CC functions, and to contaminate the CC detection of water vapor in the exoplanets.

We evaluated the noise produced in the CC functions by unocculted spots as a function of spot filling fraction, temperature, and host star spectral type, for each of the 4 molecules we study (\S\,\ref{sec: spot_noise}).  We find that filling fraction is more important than spot temperature.  Spot filling fractions as large as 10\% will degrade, but not prevent, the detections of carbon dioxide, methane, and oxygen.  Star spot water vapor produces sufficient noise to limit the detection of exoplanetary water, and it also directly contaminates the exoplanetary water detection (\S\,\ref{sec: contamination}). Correction for unocculted star spots will be needed to detect water vapor in temperate rocky planets transiting M-dwarfs. 

We describe our method of alternate normalization (\S\,\ref{sec: alternate}), that minimizes star spot effects. Although we here apply alternate normalization to simulations of temperate rocky planets transiting M-dwarfs, the concept could in principle be applied to a wide range of transit spectroscopy. We normalize the spectral flux difference (in-transit minus out-of-transit) using a proxy spectrum wherein spots have been removed.  We derive the proxy spectrum by extrapolating a series of high resolution spectra to the spotless case, using linear regression versus the star's V-K photometric color.  We demonstrate that this works in principle using PHOENIX models and plausible levels of observational random noise, and it does not depend on the nature of the star spot spectra, neither for modeled star spots nor observed ones. Nevertheless, there are challenges to be overcome, and we discuss those challenges in \S\,\ref{sec: caveats}. The challenges include: 1) possible extra cloudiness in the exoplanetary atmosphere (although we included some clouds), 2) the effect of numerous small spots that can be occulted during transit, 3) possible insufficient temporal variation of the stellar spots to enable extrapolation to the spotless case, 4) the need for caution because high resolution spectra of star spots have not yet been measured, and 5) possible dominance by faculae, and thereby the need to expand our alternate normalization method using proxies for chromospheric emission. \S\ref{sec: comprehensive} projects how some of these challenges could be overcome. If M-dwarf stellar spectra cannot be perfectly extrapolated to the spotless case, it will be helpful to directly measure the spectral properties of star spots and faculae, so as to better correct for their residual contamination. \S\ref{sec: deriving} specifically proposed a method for measuring the spectral properties of star spots and faculae using cross-correlation templates that are tuned to sets of lines having different lower state energies.

\acknowledgements

This work was performed by members of the NASA Nexus for Exoplanet System Science (NExSS) Virtual Planetary Laboratory Team, and funded via NASA Astrobiology Program grant No. 80NSSC18K0829. The work of M.H.C was additionally supported by an appointment to the NASA Postdoctoral Program at the NASA Goddard Space Flight Center, administered by Oak Ridge Associated Universities under contract with NASA.  Contributions by S.P. are supported by NASA under award number 80GSFC24M0006. We thank the referee for their critique that led us to include \S\ref{sec: comprehensive}.

% \begin{thebibliography}{references.bib}

\clearpage

\bibliography{references.bib}
\end{document}